\documentclass{jfm}
\usepackage{graphicx}
\usepackage{epstopdf, epsfig}
\usepackage{bm}
\usepackage{color}
\usepackage{amsmath}
\newcommand{\basilisk}{{\usefont{T1}{pzc}{m}{n}Basilisk} }

\shorttitle{Transient ejection from bubble bursting}
\shortauthor{A. M. Ga{\~n\'a}n-Calvo and J. M. L{\'o}pez-Herrera}

\title{On the physics of transient ejection from bubble bursting} 

\author{Alfonso M. Ga{\~n\'a}n-Calvo\aff{1,2}
  \corresp{\email{amgc@us.es}},
 \and Jos{\'e} M. L{\'o}pez-Herrera\aff{1}}

\affiliation{\aff{1}Dept. Ingenier{\'\i}a Aerospacial y Mec{\'a}nica de Fluidos, Universidad de Sevilla.\\
Camino de los Descubrimientos s/n 41092, Spain.\aff{2} Laboratory of Engineering for Energy and Environmental Sustainability,\\
Universidad de Sevilla, E-41092 Sevilla, Spain}

\begin{document}

\maketitle

\begin{abstract}
The transient ejection due to a bubble bursting at the interface of a liquid with a gas environment is here described using a dynamical scaling analysis along the process. We show here that the ejection of a liquid microjet requires the backfire of a vortex ring inside the liquid to preserve physical symmetry, which involves a non-trivial scaling. We present the first single uniformly valid expression for the size and speed of ejected droplets for the whole range of the Ohnesorge and Bond numbers where droplet ejection occurs. The evolution of the flow variables, the apparent singularity for a critical Ohnesorge number, and the dispersion of data around this point are explained. Our model generalizes or displaces other recently proposed ones, impacting for instance the statistical description of sea spray.
\end{abstract}

\section{Introduction}

Everyday experience teaches that radially convergent flows close to a liquid surface produce vigorous transient liquid ejections in the form of a jet perpendicular to the surface, as those seen after bubble bursting \citep{Kientzler1954}, droplet impact on a liquid pool \citep{Yarin2006}, or cavity collapse \citep{Ismail2018}. The mechanical energy of the flow comes from the free surface energy of the initial cavity. Among these processes, bubble bursting can be considered the parametrically simplest (just two non-dimensional parameters, v.g. the Ohnesorge (oh) and Bond (Bo) numbers determine the physics and the outcome, \cite{MG20}) and most ubiquitous one, and has received a special attention in the scientific literature due to its impact at global planetary scales (bubble bursting at the sea surface, e.g. \cite{Kientzler1954,Blanchard1957,Veron2015,Sampath2019}; disease and pandemic transmission, e.g. \cite{Bourouiba2021}) and in more indulging applications (e.g. \cite{Ghabache2016,Seon2017}).

Current works state that the subject is already deeply understood \citep{Berny2020,SLJ2021}. Yet, despite optimistic statements, even in the simplest case where the liquid properties are constant and the gas-to-liquid density and viscosity ratios are very small, unsettling but crucial questions remain open. Within an ample range of Oh and Bo, bubble bursting exhibits a strong focusing effect due to the nearly cylindrical collapse of a main wave at the bottom of the parent bubble. In these cases, a tiny bubble gets trapped and an extremely thin and rapid spout emerges. That initial spout grows into a much taller, thicker and slower transient jet that may eventually expel droplets. Their eventual size and speed is determined by Oh and Bo. However, for a critical Oh, the initial spout keeps its large speed and extremely thin radius until it ejects a nearly invisible droplet, such that the spout momentum (proportional to its velocity times its volume) vanishes. Strikingly, a nearly symmetric behavior is observed around that apparently critical Oh (\cite{Seon2017}, fig. 16). The possible existence of a mathematical singularity that may explain this phenomenon is dismissed by many. In reality, around that special parametrical region, compressibility and possibly nonlinear surface effects set in and the experiments exhibit a strong dispersion that may be reduced relying to local viscous effects and interaction with the gas environment \citep{Brasz2018,Dasouqi2021}. Appealing to the facts, the global streamlines of the flow resemble a dipole with its axis perpendicular to the interface (figure \ref{fig1}). In general, the energy excess of the convergent flow leads to {\sl both} a fast transient {\sl capillary jet} and a {\sl vortex ring} in opposite directions but equivalent effective momenta by virtue of Newton's third law of motion. Their radically different kinematics, due to a large density disparity across the interface, is not an obstacle to their profound symmetry, as we will show here. This symmetry and its scalings are the keys to determine the eventual ejected droplet size and speed, and the appearance of singular dynamics. These crucial observations cannot be accommodated by recent proposals \citep{Gordillo2019}.

In this work, we present: (i) A universal non-trivial scaling of the evolution of the flow kinematic variables (characteristic lengths and velocity), (ii) a prediction of the eventual ejected droplet size and speed, for the complete range of Oh for which ejection occurs (which generalizes \cite{G17}), and  (ii) the existence of an apparently singular value of Oh explaining the physical observations. Our theoretical results are carefully validated against existing data from prior works (e.g. those collected in \cite{G17,Brasz2018,GananCalvo2018,Deike2018,Berny2020}) and novel numerical simulations.

\section{Formulation and dynamical scaling analysis}
\label{BB}

Consider a gas bubble of initial radius $R_o$ tangent to the free surface of a liquid with density, surface tension and viscosity $\rho$, $\sigma$ and $\mu$, respectively, in static equilibrium under a relatively small action of gravity acceleration $g$ normal to the free surface far from the bubble. The flow properties are considered constant. At a certain instant, the thin film at the point of tangency breaks and the process of bubble bursting starts. The problem is thus determined by the Ohnesorge (Oh$ = \frac{\mu}{(\rho\sigma R_o)^{1/2}}$) and Bond (Bo$ = \frac{\rho g Ro^2}{\sigma}$) numbers alone. The liquid rim that is initially formed pilots two main capillary wave fronts, one that advances along the bubble surface towards its bottom, and the other propagating away from the cavity. In some parametrical regions of the domain $\{$Bo, Oh$\}$ of interest, those main waves may form wavelets at their front as those produced by the capillary rollers and bores described by \cite{LH92}. These wavelets may arrive first to the bottom, but do not produce any significant effect compared to the collapsing main wave \citep{GananCalvo2018} in the parametrical domain here considered. When the main wave front approaches the bottom, it becomes steep. If the wave front leads to a nearly cylindrical collapsing neck in a certain region, a tiny bubble gets trapped below the surface once the ejection process commences. This process of radial collapse and bubble trapping is locally described by the theory of \cite{EFLS2007}, \cite{FSE11}, and \cite{EF15} when a strong asymmetry in the axial direction occurs. This asymmetry is precisely the responsible of the initial extremely rapid and thin ejection in the opposite direction from the trapped bubble.

\begin{figure}
	\centering
	\includegraphics[width=0.95\textwidth]{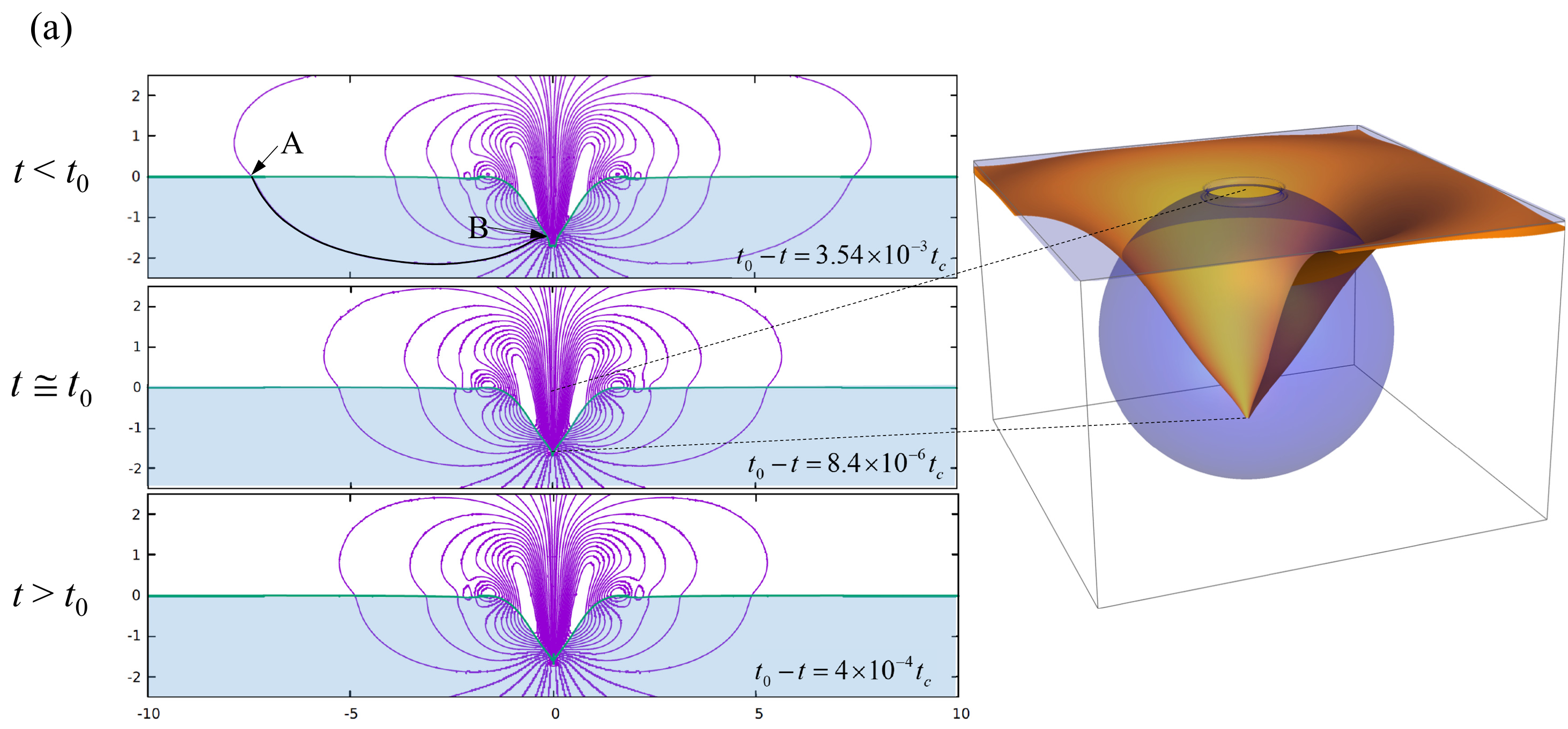}
    \includegraphics[width=0.95\textwidth]{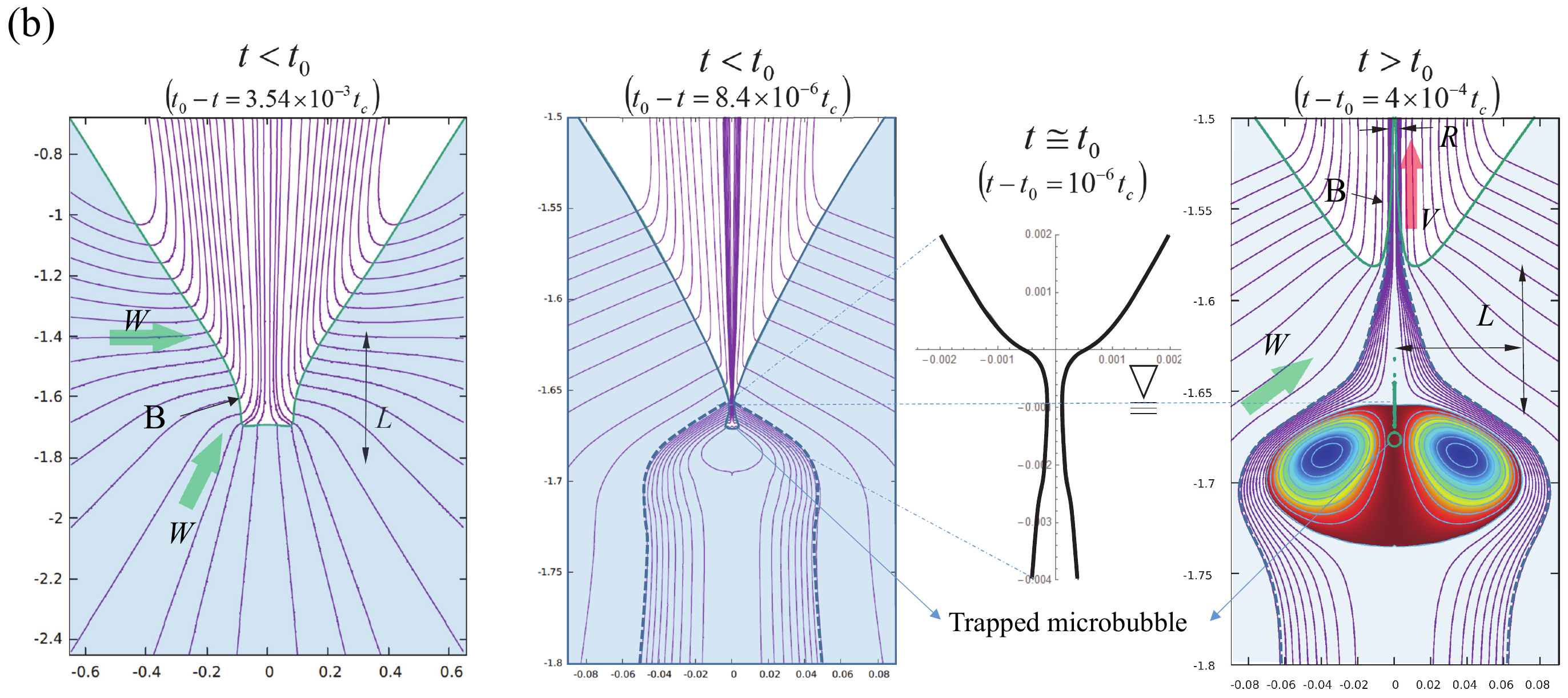}
	\vspace{-3mm}
	\caption{General overview of the flow development around the critical time $t_0$ of collapse of the main pilot wave at the bottom of the cavity, for Oh = 0.032, Bo = 0. The three instants here illustrated are $t_0-t=3.54\times 10^{-3}tc$, $t_0-t=8.4\times 10^{-6}tc$, and $t-t_0=4\times 10^{-4}tc$: (a) Global flow streamlines (similar to dipole contours at a  liquid-gas surface) showing a particular one (line A-B) ending at the point (B) just above where the collapse eventually occurs. The blue sphere of the 3D rendering is the initial bubble immediately after bursting. (b) Local details of the same instants. The stream function levels showed are closer around the tiny trapped bubble to exhibit the vortex ring. Noteworthy, the upper axial point of the ellipsoid defining the vortex ring coincides with the point where the surface collapsed at $t=t_0$ and remains at that position: observe the horizontal line connecting the three panels of figure (b). The main flow velocities $W$ (radial) and $V$ (axial, ejection) are indicated. The dashed lines represent the instantaneous stream tube which feeds the collapsing point or the issuing jet. In these simulations using VOF (\basilisk, \cite{Basilisk}), the density and viscosity of the liquid is 1000 and 100 times that of the gas, respectively, to reflect similar relations to the air-salt water ones.}
	\label{fig1}\vspace{-2mm}
\end{figure}

Figure \ref{fig1} shows three illustrating instants of the flow development around the critical instant $t_0$ at which the axial speed of the liquid surface on the axis reaches its maximum (immediately after the tiny gas neck collapses). When $t<t_0$, the flow is predominantly radial with a characteristic speed $W$ at the interface. After collapse ($t>t_0$), while the main flow keeps running radially with speed $W$, the axial asymmetry of the flow and the kinetic energy excess of the liquid is axially diverted in the two opposite directions, with radically different results: (i) towards the open gas volume, producing the liquid spout with speed $V$ and characteristic radial length $R$; and (ii) towards the liquid bulk, as a backward reaction vortex ring (fig. \ref{fig1}b, in colors) with characteristic length $L$, around the trapped microbubble. Eventually, the advancing front of the resulting capillary jet expels a droplet or droplet train scaling as $R$.

Obviously, all scales $\{W,V,R,L\}$ are time-dependent. Prior works (e.g. \cite{ZKFL00,Lai2018}) have focused on time self-similarity of flow variables, arriving to fundamental universalities characterizing their evolution. Assuming that those time self similarities exist, our focus is here to obtain closed relationships among those scales and to predict the size and speed of ejections reflected by the eventual values of $R$ and $V$ at the end of the process. This approach, though, demands the identification of the key symmetries appearing in the problem that may lead to an effective problem closure \citep{GRM13}. To this end, in Ga{\~n\'a}n-Calvo (2017, 2018) a set of relations among the radial and axial characteristic lengths and velocities is formulated on the basis that inertia, surface tension and viscous forces should be comparable very close to the instant of collapse of the free surface. While the previously proposed simple relations were fundamentally consistent, a rigorous derivation is here offered under the light of detailed numerical simulations. We will show that the formulated scale relationships not only hold at specific instants of their development, but throughout the entire process, for all Oh values of interest.

\subsection{An integral relations-based dimensional analysis}

The momentum equation of the liquid can be written as:
\begin{equation}
\rho \mathbf{v}_t+ \nabla\left(\rho \mathbf{v}^2 /2 + p - P_a + \rho g z\right)=\mathbf{v}\wedge\nabla\wedge \mathbf{v}+\mu \nabla^2 \mathbf{v}.
\label{mom1}
\end{equation}
where $\mathbf{v}$ is the velocity vector, subindex $t$ denotes partial derivative with time, $p$ is the liquid pressure, $z$ the axial coordinate, $\mathbf{n}$ the unit normal on the liquid surface, and $P_a$ the static gas pressure. Equation (\ref{mom1}) can be multiplied by the unit vector ${\boldsymbol l}$ tangent to any instantaneous streamline, in particular the streamline flowing through a point A where it meets the free surface to a point B at the vicinity of the point of collapse (see figure \ref{fig1}(a)). Integrating with respect to the streamline coordinate $s$ from A to B yields:
\begin{equation}
\rho \int_A^B {\boldsymbol l}\cdot \mathbf{v}_t \text{d}s + \left. \rho \mathbf{v}^2 /2\right|_B + \left.\sigma \nabla\cdot \mathbf{n}\right|_B
 +\left. \rho g \Delta z\right|_A^B=\mu \int_A^B {\boldsymbol l}\cdot \nabla^2 \mathbf{v}\text{d}s,
\label{momi}
\end{equation}
since the velocity is negligible at A, and pressure is $P_a$. $\Delta z$ is the depth of point B respect to A. As a general consideration, the liquid velocities are very small everywhere compared to the velocity at distances $L$ to the collapsing region, which may exhibit a self-similar flow structure \citep{ZKFL00, Duchemin2002,Lai2018}. The length scale $L$ also characterizes the inverse of the mean local curvature of the liquid surface around the region of collapse, for any given time $t$. Incidentally, $L$ is also the scaling of the boundary layer due to the main wave \citep{GananCalvo2018}, supporting the central argument in \cite{G17}. Thus, $L$ obviously changes with time around the instant of collapse. To look for symmetries around $t_0$, let us consider two situations, one for $t<t_0$ and the other for $t>t_0$ such that their characteristic length scales $L$ are {\sl the same} (see figure \ref{fig1}). Then, one may estimate the characteristic values of each term of equation (\ref{momi}) for both $t<t_0$ and $t>t_0$. In any case, the first two terms of (\ref{momi}) are always of the same order for this unsteady problem.

For $t<t_0$, both the left integral and the kinetic energy term at B in (\ref{momi}) should scale as $\rho W^2$. The surface tension term at B should be proportional to $\sigma/L$, and the gravity term to $\rho g R_o$. Finally, the curvature of streamlines suggests that the viscous stresses should scale as $\mu W/L$ (boundary layers cannot be thinner than $L$, as shown by \cite{GananCalvo2018}), as it does the integral in the right hand side of (\ref{momi}). Thus, the overall scaling balance of (\ref{momi}) can be expressed as:
\begin{equation}
\rho\left(W^2+\beta_1 g\,R_o \right)\sim \sigma\, L^{-1}+\alpha_1 \mu\, W\, L^{-1},
\end{equation}
where prefactors $\alpha_1$ and $\beta_1$ should be universal constants reflecting the dominance or sub-dominance of the corresponding terms. Using the natural length and velocity, $l_\mu=\mu^2/(\rho \sigma)$ and $v_\mu=\sigma/\mu$ respectively, and defining $\zeta=L/l_\mu$, $\omega=W/v_\mu=$ and $\epsilon_1=\beta_1 \text{Oh}^2\text{Bo}$, this equation can be written as:
\begin{equation}
\omega^2+\epsilon_1 \sim \left(1+\alpha_1 \omega \right)\zeta^{-1}.
\label{e1}
\end{equation}
When $t>t_0$, the vigorous ejection in the axial direction 
generates the new characteristic velocity $V$ and radial size $R$ scales of the liquid jet. To appraise physical symmetry, the scale of the radial velocity $W$ in the bulk and the (new) axial scale of the jet $L$ should be taken the same as their namesakes for $t<t_0$ (see central panel in figure \ref{fig1}(b) for Oh = 0.032, Bo = 0). Even before collapse, the flow develops a stagnant region in the liquid \citep{G17} just below the collapsing point. From this point, a vortex ring of characteristic (growing) size $L$ and speed $W$ like those beautifully described by \cite{Maxworthy1972} develops: see figure 4 of his work (dissecting absolute and relative speeds is not trivial in these flows, as Maxworthy showed). Assuming that the trapped microbubble is much smaller than the vortex, by virtue of total energy preservation the mechanical energy of the vortex, $\rho W^2L^3$, should mirror that of the vigorous microjet, $\rho V^2 R^2 L$. The kinematic proposal of \cite{Gordillo2019} is inconsistent with this comprehensible mechanical symmetry principle, subsequently demonstrated by numerical simulation: if one observes the streamlines below the jet, it is completely fed by an axial stream coming from the utmost depths and surrounds the vortex ring (see figure \ref{fig1}b, right panel). Those streamlines should ultimately originate very far from the bubble at the free surface. In summary, one can write:
    \begin{equation}
    \rho V^2 R^2 L \sim \rho W^2 L^3\Longrightarrow V\,R\sim W\,L\Longrightarrow \upsilon\, \chi\sim \omega\, \zeta
    \label{e2}
    \end{equation}
where $\chi=R/l_\mu$ and $\upsilon=V/v_\mu$. As a secondary consequence, the nearly conical surface {\sl independently} raises at characteristic speed $W$ due to the incoming nearly conical flow.

One can now estimate the scaling of the different terms of (\ref{momi}) for a streamline ending at a point B at the surface of the jet (see figure \ref{fig1}a), and write the following balance:
\begin{equation}
\rho\left(V^2+\beta_2 g\,R_o \right)\sim \sigma\, R^{-1} + \alpha_2 \mu W\, L^{-1},
\label{t2}
\end{equation}
where, again, prefactors $\alpha_2$ and $\beta_2$ should be universal constants at the end of the process. 
Using natural scales and defining $\epsilon_2=\beta_2 \text{Oh}^2\text{Bo}$, (\ref{t2}) reads:
\begin{equation}
\upsilon^2+\epsilon_2 \sim \chi^{-1}+ \alpha_2 \omega \zeta^{-1},
\label{e3}
\end{equation}

As in \cite{G17}, equations (\ref{e1}), (\ref{e2}) and (\ref{e3}) define three explicit relationships among the time dependent scales $\{\chi,\upsilon,\zeta,\omega\}$. For instance, defining
\begin{align}
\varphi & = \alpha_1/2 + \left((\alpha_1/2)^2+\zeta(1-\epsilon_1 \zeta)\right)^{1/2},\nonumber \\
\phi & = 1/2+\left(1/2+\varphi^2\left(\alpha_2\varphi/\zeta^2-\epsilon_2\right)\right)^{1/2},
\label{def1}
\end{align}
one obtains the following explicit relationships:
\begin{equation}
\chi \sim \varphi^2 \phi^{-1},\,\, \upsilon \sim \varphi^{-1}\phi,\,\, \omega \sim \varphi/\zeta.
\label{ps1}
\end{equation}
For $\alpha_1\rightarrow 0$, $\alpha_2\gg 1$ and $\epsilon_{1,2}=0$, (\ref{ps1}) yields $\chi \sim \upsilon^{-5/3}$, $\zeta \sim \upsilon^{-4/3}$ and $\omega \sim \upsilon^{2/3}$, as predicted by \cite{G17}. However, given that the ejection is fundamentally a ballistic (inertial) process, one may expect two possible regimes in the evolution of the ejection: (1) When $t-t_0<t_\mu=\mu^3/(\rho \sigma^2)$, the inertia and viscous forces should be dominant. In this case, for Bo = 0, $\chi \sim \zeta \sim \upsilon^{-2}\sim\omega^{-2}$, consistently with \cite{ZKFL00}. (2) For $t-t_o>t_\mu$, inertia and surface tension forces should prevail. Here, if Bo = 0, one obtains $\chi\sim \zeta \sim \upsilon^{-1} \sim \omega^{-1}$.

\begin{figure}
	\centering\includegraphics[width=0.55\textwidth]{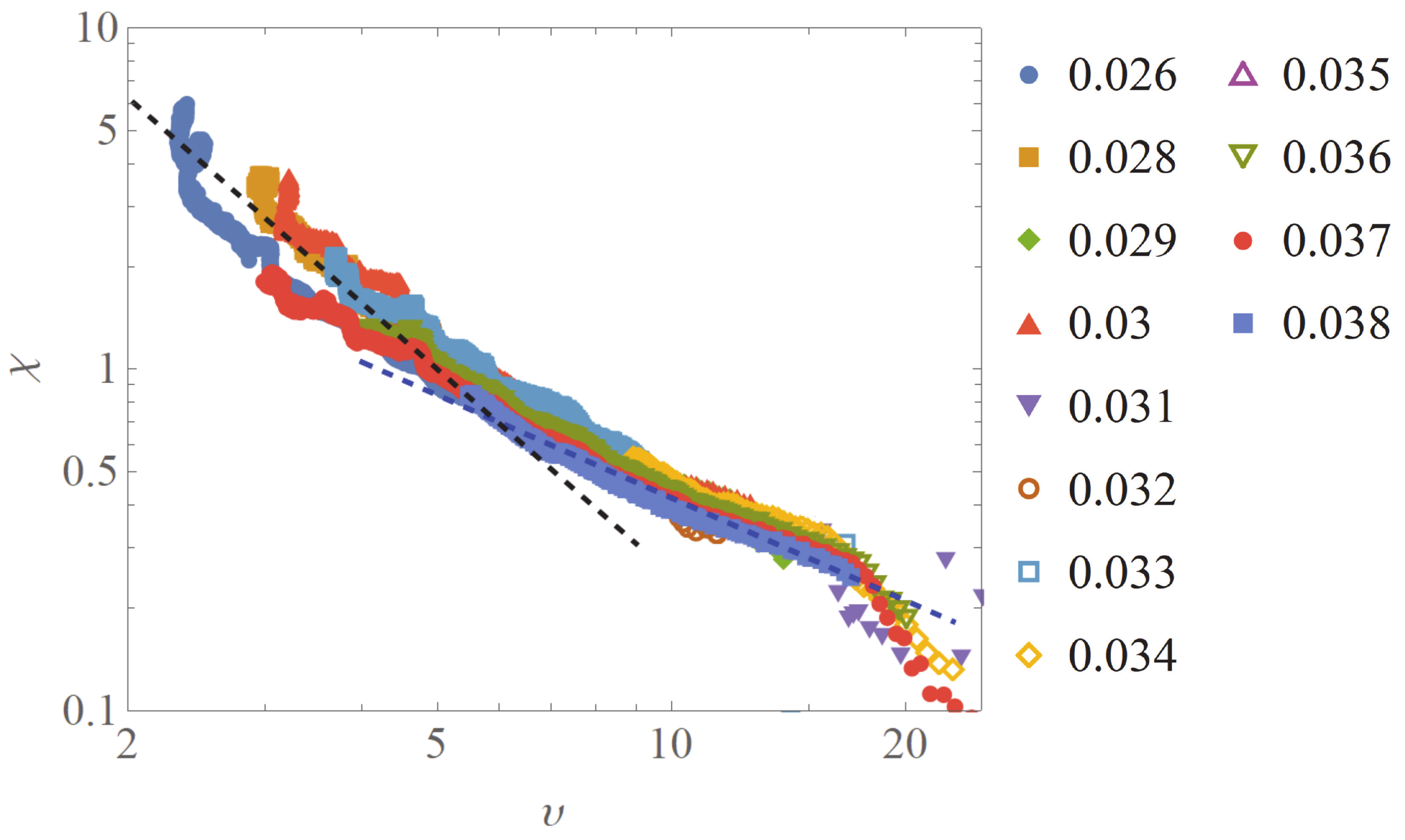}
	\vspace{-3mm}
	\caption{Numerical data (momentum-conserving VOF, \basilisk, minimum cell size $3\times 10^{-4}R_o$ for this figure) compared to prediction (\ref{ps1}) for $\chi$ vs $\upsilon$. The black dashed line is the theoretical prediction for $(t-t_o)>t_\mu$ ($\alpha_{1,2}\rightarrow 0$), which yields $\chi \sim \upsilon^{-2}$. The blue dashed line would correspond to $0<(t-t_o)<t_\mu$, with $\alpha_{1,2}\gg 1$, yielding $\chi \sim \upsilon^{-1}$. Observe the visible change of regime in the case Oh = 0.038 around $\chi\simeq 0.65$, i.e. around $R \simeq 0.65 l_\mu$.}
	\label{fig2}\vspace{-2mm}
\end{figure}

To assess the validity of (\ref{ps1}), we analyze the time-dependent evolution of the scaling variables $\{\chi,\zeta,\upsilon,\omega\}$ obtained from numerical simulation (momentum-conserving VOF, \basilisk). We have recorded the time-dependent values of the radius of curvature and velocity of the jet front at the jet axis to represent the characteristic time-dependent values of $R$ and $V$, respectively, for $t>t_0$ (no scales of $R$ and $V$ appear for $t<t_0$). Since (\ref{ps1}) does not provide the time-dependencies of the variables, in figure \ref{fig2} we plot the non-dimensional evolving values of $\chi$ vs $\upsilon$ for a detailed range of Oh numbers. Firstly, the collapse is remarkable. Secondly, we observe the two clear regimes as anticipated: viscosity-dominated ($\alpha_{1,2}\gg 1$) when $\chi \lesssim 1$ ($R \lesssim l_\mu$) and surface tension-dominated ($\alpha_{1,2}\rightarrow 0$) when $\chi \gtrsim 1$ ($R \gtrsim l_\mu$), corresponding to $\chi\sim \upsilon^{-2}$ and $\chi \sim \upsilon^{-1}$ respectively.
The formation and pinch-off of a droplet is expected to take place once surface tension becomes important, resulting in a dependency as observed. The relatively small deviations from the theoretical prediction (\ref{ps1}) for the smallest and largest values of $\upsilon$  reflect the vibrations of the droplet about pinch-off and ejection (not considered here) and the very initial evolution just after $t=t_0$, respectively.

\subsubsection{Problem closure: ejected droplet size and speed}

To close the problem, a key question which is not resolved by (\ref{ps1}) is the mechanical energy available to perform the ballistic ejection close to $t_o$. One can express the mechanical energy equation in a sufficiently ample fluid volume $\Omega(t)$ around the initial bubble from the instant of bubble bursting ($t=0$) up to the point of collapse $t_o$ when the minimum scale is $l_\mu$. To do so, the fluid volume $\Omega(t)$ can be a hemisphere around the cavity with a radius about twice or three times larger than $R_o$:
\begin{equation}
\left. \int_{\Omega(t')}\rho\left(\mathbf{v}^2/2+gz\right)\text{d} \Omega \right|_{t=0}^{t}=-\int_{t=0}^{t} \int_{S(t')} \mathbf{v}\cdot \left(  {\boldsymbol\tau'} -p\mathbf{I} \right) \cdot \mathbf{n}\, \text{d} A\, \text{d} t'
\label{inte}
\end{equation}
where ${\boldsymbol\tau'}$ is the viscous stress and $\mathbf{I}$ the identity matrix. We define the capillary velocity as $V_o = \left(\frac{\sigma}{\rho R_o}\right)^{1/2}$, which is the global expected value of the average velocities along the process. Observing the geometric configuration of streamlines in figure \ref{fig1}, the most relevant feature is a stream tube with velocity $W$, transversal area proportional to $L^2$ and total length to $R_o$ which ends up at the point of collapse (figure \ref{fig2}b). Thus, the kinetic energy term should be proportional to $\rho W^2 L^2 R_o$ (the initial kinetic energy is zero). If one neglects viscous and gravity works, expression (\ref{inte})suggests the scaling $\rho W^2 L^2 R_o \sim \sigma R_o^2 \sim \rho V_o^2 R_o^3 = $ const.. This relationship was already observed by \cite{Duchemin2002} and \cite{Ghabache2014}, among others, for negligible viscosity and gravity. However, it does not faithfully describe the nonlinear behavior observed for non-zero Oh, especially around Oh $\simeq 0.033$ (e.g. \cite{Duchemin2002,Walls2015,Ghabache2016a,Seon2017,G17,Deike2018}), where the product $V R$ becomes much smaller than $V_o R_o$. To capture that behavior one should retain gravity and viscous terms: While the former is simply volumetric and proportional to $(\rho g R_o)R_o^3$, the latter should be proportional to the dominant one, i.e. $(\mu V_o/R_o)R_o^2\times R_o$, which should naturally appear as an energy sink with its corresponding sign. Note that the average total displacement of fluid particles should be proportional to $R_o$. However, a nearly perfect balance between the global surface tension and the viscosity terms occurs around Oh $\simeq 0.033$, with little excess for ejection left. This leads to the ballistic emission of an extremely thin ligament of vanishing radial size and very large speed. During that extreme emission, while $\mu V/R$ stays comparable to $\rho V^2$, viscous forces decelerate the spout until $\sigma/R$ becomes strong enough to form a droplet at its tip. Hence, the size of that minimum drop should reflect the ultimate balance between inertia, viscosity and surface tension forces, i.e. $R \sim l_\mu$. Thus, a {\sl minimum} surface tension energy proportional to $(\sigma/l_\mu) l_\mu^2\times R_o$ should necessarily remain available along the process for the emission and formation of that droplet, in consistency with experimental observations. In conclusion, keeping that minimum in mind, one can formulate the following scaling of (\ref{inte}):
\begin{equation}
\rho \left(W^2 L^2 + \beta_3 g R_o^3\right) R_o \sim \sigma R_o^2\left({\text{Oh}^*}^2 + l_\mu/Ro\right) - \alpha'_3 \mu V_o R_o^2
\label{e3e}
\end{equation}
where, again, prefactors Oh$^*$, $\alpha'_3$ and $\beta_3$ should be universal positive constants for very small gas-to-liquid density and viscosity ratios. (\ref{e3e}) is equivalent to the energy equation in \cite{G17}, with the exception of the {\sl minimum} contribution of the surface tension at the scale $l_\mu$. In terms of the natural scales, (\ref{e3e}) writes
\begin{equation}
\upsilon^2 \chi^2 \sim \left(\text{Oh}^*/\text{Oh} - 1\right)^2+\alpha_3/\text{Oh} - \beta_3\text{Oh}^{-2}\text{Bo}\equiv \xi - \beta_3\text{Oh}^{-2}\text{Bo},
\label{Solution}
\end{equation}
where $\alpha_3 \equiv 2$ Oh$^*-\alpha'_3$ and $\xi \equiv \left(\text{Oh}^*/\text{Oh} - 1\right)^2+\alpha_3/\text{Oh}$. The formal structure of (\ref{Solution}) aims to capture the {\sl almost} perfect hypothesized balance at Oh$^*$ in the absence of gravity. In effect, from (\ref{Solution}) the value of Oh where $R$ ($V$) is minimum (maximum) is given by:
\begin{equation}
\text{Oh}_c=\frac{{\text{Oh}^*}^2-\beta_3\text{Bo}}{\text{Oh}^*-\alpha_3/2}\equiv \frac{2({\text{Oh}^*}^2-\beta_3\text{Bo})}{\alpha'_3},
\label{Ohc}
\end{equation}
around which the dependence is nearly symmetric. When Bo = 0, if $\alpha_3$ is sufficiently small, Oh$_c\simeq $ Oh$^*$. As we will see next, experiments reveal that $\alpha_3\ll $ Oh$^*$, confirming our hypothesis. This fundamental finding explains and justifies the experimental observations by many authors (\cite{Duchemin2002,Walls2015,Seon2017,Ghabache2016a,Brasz2018,Deike2018,Berny2020}, among others), but not only this: experiments confirm the {\sl continuous} validity of (\ref{Solution}) for the whole Oh range where droplet ejection occurs.

\subsection{Experimental verification}

(\ref{ps1}) and (\ref{Solution}) yield the scaling of $\{\upsilon,\omega,\chi,\zeta\}$ as functions of Oh and Bo. The seven universal constants $\{\text{Oh}^*,\alpha_{i=1,2,3},\beta_{i=1,2,3}\}$ will be obtained from experiments, which show that $\alpha_{1,2}\rightarrow 0$ is a very good approximation, consistently with the fact that surface tension should become dominant when droplet formation and ejection occurs. To verify our model and obtain the relevant scaling constants for the whole Oh range experimentally explored, six hundred experimental and numerical measurements of first ejected droplets and about one hundred of their corresponding initial velocities have been collected from the literature. In addition, to appraise the predictive value of (\ref{Solution}) in the range Oh$\in (0.026,0.05)$, where the maximum variation among the published results is found, additional extensive numerical simulations using a momentum-conserving VOF scheme (\basilisk) have been performed. Simulations are made with a gas-to liquid viscosity and density ratios equal to 0.01 and 0.001, respectively, up to a minimum cell size $1.5\times 10^{-4}R_o$ around the point of collapse and ejection, equivalent to about 2 nm for a gas bubble in water with Oh = 0.033.

\begin{figure}
	\centering\includegraphics[width=0.95\textwidth]{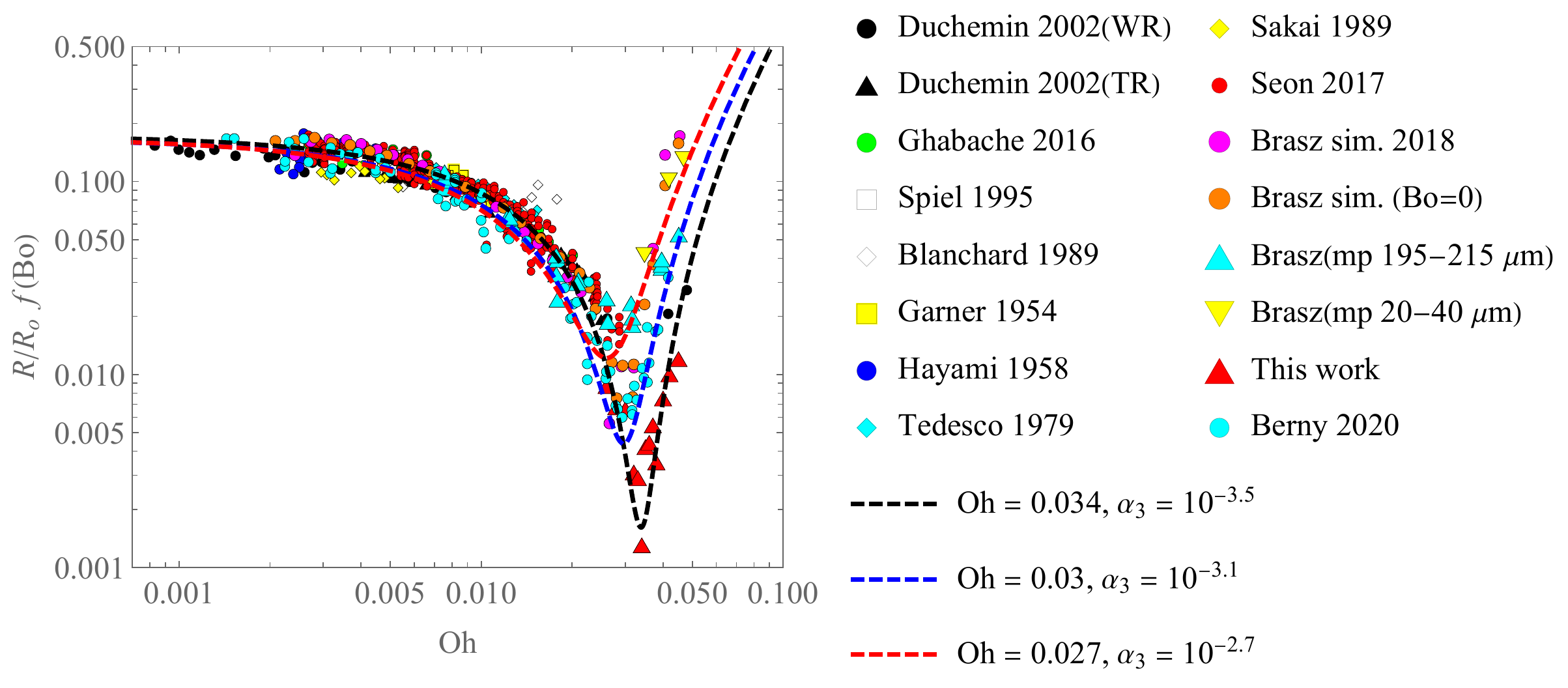}\\
\centering\includegraphics[width=0.78\textwidth]{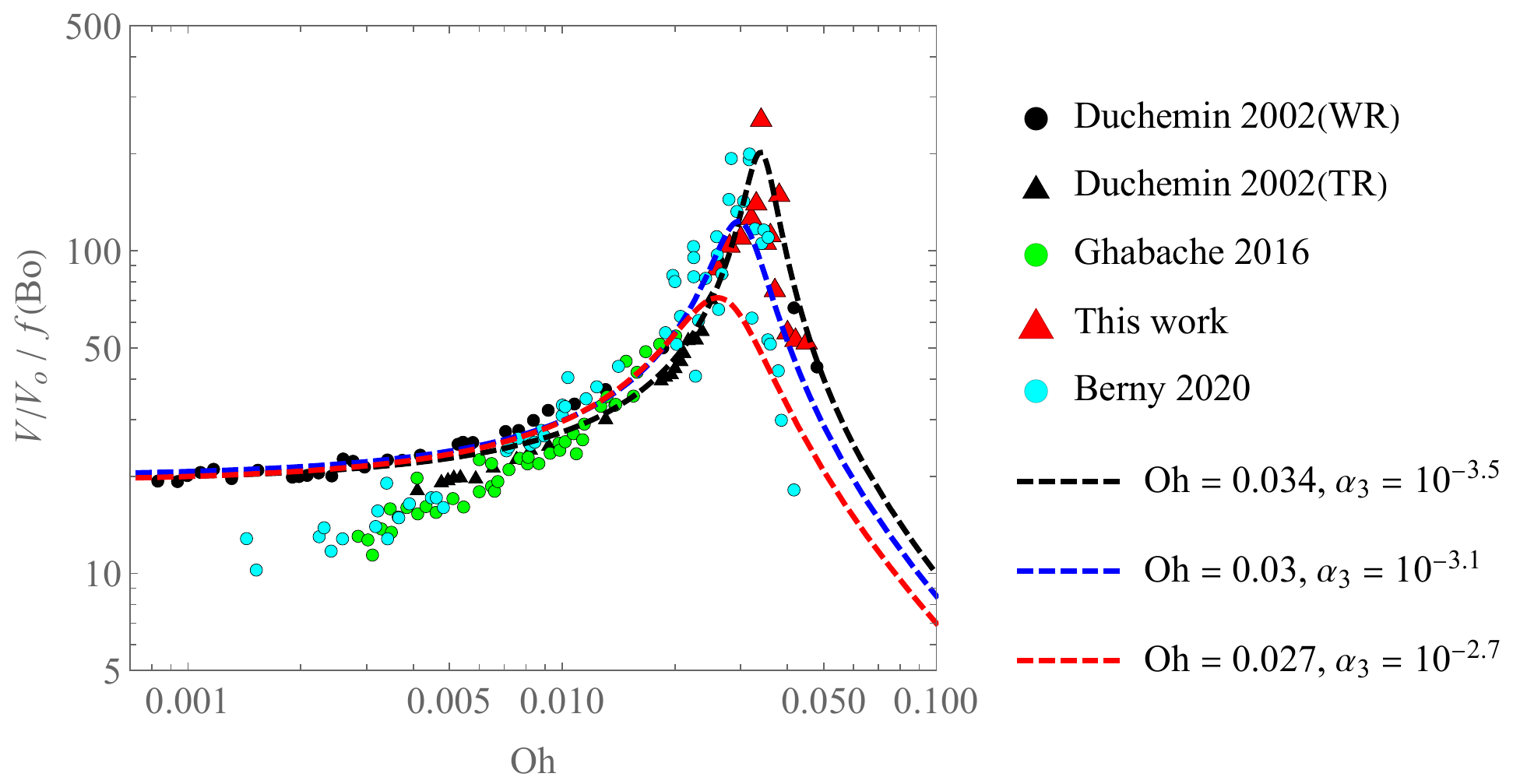}
	\vspace{-3mm}
	\caption{Experimental measurements from different literature sources (see \cite{G17} for additional information). The plots show the first droplet size $R$ (a) and the initial speed of ejection $V$ (b) represented by $R/R_o=$ Oh$^2\chi$ and by $V/V_o=$ Oh$^{-1}\upsilon$, corrected with the function $f$. Dashed lines give the theoretical predictions from present model. The Oh range covers seawater bubbles in the range from 8 $\mu$m to 2 mm.}
	\label{fig3}\vspace{-2mm}
\end{figure}

The first ejected droplet radius $R$ and its speed $V$ data, expressed in non-dimensional form as $\chi \cdot f$ and $\upsilon/f$ at the time of ejection, respectively, are plotted in figure \ref{fig3}(a) and (b). From (\ref{ps1}) and (\ref{Solution}) with $\alpha_{1,2}=0$, the correction factor $f$ is:
\begin{equation}
f=\left(1-\beta_3\text{Bo}/\xi\right)^{-1}\phi,
\end{equation}
where the values of $\phi$ and $\xi$ are calculated from the experimental data and using the fitting constants $\alpha_3$ and $\beta_{1,2}$.

The dashed lines are the theoretical predictions assuming constant liquid properties $\{\rho,\sigma,\mu\}$. First, for $\alpha_{1,2}\lesssim 10^{-2}$ the fitting remains unaltered, and thus $\alpha_{1,2}=0$ is assumed for simplicity. If we use our numerical simulations for Oh $\in (0.026,0.05)$ and the rest of data for Oh $< 0.028$, the optimum fitting by least squares yields Oh$^*=0.034$ and $\alpha_3=10^{-3.5}$. However, a fine fitting to other experimental data series is possible for slightly different values of Oh$^*$ and $\alpha_3$ (see figure \ref{fig3}), which reflects the extreme sensitivity of the balance at Oh = Oh$^*$ in the Oh range between 0.025 and 0.035. All this confirms our hypothesis and the continuous validity of (\ref{Solution}) for the whole Oh range.

The experiments also give a best collapse for $\beta_2 {\text{Oh}^*}^2\simeq 0.16$ and $\beta_3\simeq 0$. However, since different authors usually measure the ejection speed when the droplet reaches the free surface, not just at the point of ejection \citep{GananCalvo2018}, separate fittings yield $\beta_3 \simeq 5\times 10^{-4}$ for the droplet size and $\beta_3 \simeq -2\times 10^{-4}$ for the ejection speed. Besides, the deviation observed in the data for the smallest Oh range from both \cite{Ghabache2016a} and \cite{Berny2020} evinces the effect of the gravity on the initial shape of the bubble, which is not considered in this work (observe that when Bo $=0$, numerical results from \cite{Duchemin2002} and our model are coincident).

\subsubsection{The critical Ohnesorge number. Resolving the paradox and the statistical nature of bursting around Oh$^*$}

A striking lack of agreement on the issue of the critical Oh has already been noticed in recent publications \citep{Walls2015,Brasz2018,Deike2018,G17,Berny2020}. As anticipated, the large data dispersion observed around Oh$_c$ (negligible Bo in all those cases) is consistent with the difficulty of measurements and simulations around these Oh values, where the slightest effect (real or numerical) produces large deviations. This includes the interaction with the gas environment, gas compressibility, but fundamentally surface effects (presence of contaminants, surfactants, particles, etc.). Owing to the inherently statistical nature of these effects, a solid ground to quantitatively represent their impact on the statistical distribution of sea spray, for instance, was needed. Although this work is focused on the first issued droplet (\cite{Berny2020} present a detailed analysis of all issued droplets from the scaling analysis of the first), we show here that just two constants, v.g. Oh$_c$ and $\alpha_3$ are enough to reflect the critical balance described around the point of collapse and to capture the entire physics of the process. Of course, the complexity of the phenomenon can be augmented including surface viscosity \citep{Ponce-Torres2017}, Marangoni and non-Newtonian effects \citep{SLJ2021}, or the presence of immiscible liquids \citep{Yang2020}, but these are topics of subsequent studies in light of current results.

This research has been supported by the Spanish Ministry of Economy, Industry and Competitiveness under Grants DPI2016-78887 and PID2019-108278RB, and by Junta de Andaluc\'{\i}a under Grant P18-FR-3623.


\end{document}